# OpenCL 2.0 for FPGAs using OCLAcc

Franz Richter-Gottfried and Alexander Ditter and Dietmar Fey
Chair of Computer Science 3 (Computer Architecture)
Friedrich-Alexander-Universität Erlangen-Nürnberg (FAU)
91058 Erlangen, Germany
Email: {franz.richter-gottfried, alexander.ditter, dietmar.fey}@fau.de

*Abstract*—Designing hardware is a time-consuming and complex process. Realization of both, embedded and high-performance applications can benefit from a design process on a higher level of abstraction. This helps to reduce development time and allows to iteratively test and optimize the hardware design during development, as common in software development. We present our tool, OCLAcc, which allows the generation of entire FPGA-based hardware accelerators from OpenCL and discuss the major novelties of OpenCL 2.0 and how they can be realized in hardware using OCLAcc.

FPGA, OpenCL, High level synthesis, High performance computing

## I. INTRODUCTION

Focusing on embedded systems typically offers two ways to realize the core components: (i) writing software for pre-built devices or (ii) designing custom hardware for the task. Consequently, this decision is usually made very early during development. As it is not easy to find an optimal solution before having designed algorithms and knowing their specific requirements, Hardware/Software Codesign is delaying this decision as long as possible. But still, the individual hardware and software components are developed in mostly disjoint design flows. The most crucial difference between software development and hardware design is the way resources are managed by the developer. While software development assumes a particular hardware to run on and perform best if memory bandwidth and the CPU utilization are saturated, e.g., by exploiting caches or vector instructions; the number of degrees of freedom is even higher when taking the hardware design into consideration, where computational resources and parts of the memory can be tailored to the application's needs.

With the increasing level of parallelism in current processors, language extensions have been developed with the goal to effectively and efficiently distribute work and data on a more abstract and thus, more device independent level to ease development, optimize the solution and increase reusability. These extensions include vendor-specific languages, e.g., CUDA, but also the vendor and device independent OpenCL-standard. Altera was the first FPGA vendor to offer a complete toolchain to implement applications on FPGAs only using OpenCL [1], by deriving an application specific data path form the high level description. Xilinx later also integrated an OpenCL fronted in Vivado HLS and recently presented SDAccel [2]. Though these approaches look very comfortable and promising, performance and efficiency is often limited [3]. In contrast, SOpenCL [4] uses a fixed data-path on the FPGA, which we think leads to a less efficient solution because the application's peculiarities, e.g., the width of data, cannot be exploited. Like the solutions of Altera and Xilinx, our approach derives the FPGA design

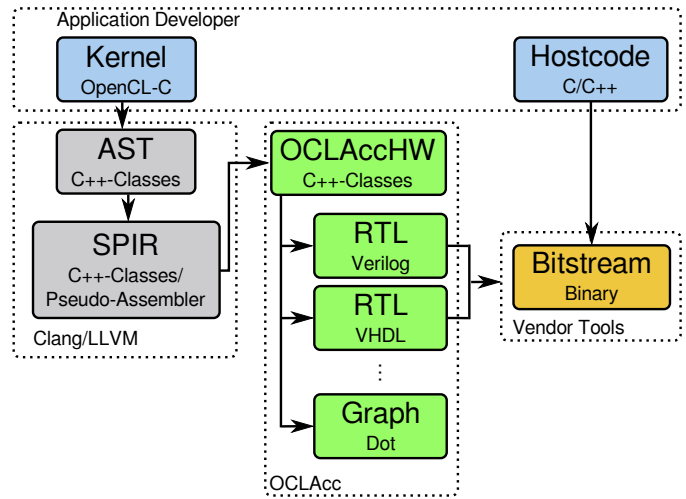

Fig. 1: Overview

directly from the algorithm and thus, allows optimal power- and time-efficiency.

In this paper we present OCLAcc [5], an open source generator for configurable logic block based accelerators, and discuss how recent extensions to the OpenCL standard can be applied to hardware design. First, the basic idea of OpenCL is presented in Section II. Section III introduces OCLAcc. New features of OpenCL are discussed in Section IV, followed by the conclusion.

## II. OPENCL

OpenCL is a hardware independent framework to separate administration of a program and computation. Administration includes, but is not limited to, coordination of computation, memory allocation or communication, all done by host-code, usually executed by a normal CPU. Computation, expressed as kernel, runs on accelerators like GPUs, but also normal on CPUs. The main advantage of OpenCL is that programmers explicitly parallelize their algorithms. However, portable code does not guarantee portable performance. To fully exploit a target's performance, its properties have to be taken into account, but this is done by methods developers are familiar with, like loop unrolling or latency hiding. For this reason we think OpenCL offers a promising way for software developers to create FPGA-based accelerators.

## III. OCLACC

Figure 1 shows the translation process of OCLAcc. The OpenCL kernel, written in a C-dialect, is translated to Standard Portable Intermediate Representation SPIR by a modified version of Clang, maintained by Khronos. SPIR itself is based on LLVM-IR, so we integrated OCLAcc into LLVM. Translation in OCLAcc happens in two steps. First, SPIR is used to generate OCLAccHW, an internal representation of the data flow, optimized to derive hardware from it. This step is based on basic blocks, which are instruction sequences of maximal length with a single entry and exit in the control flow of the kernel. Inputs and outputs are analyzed to identify streams from and to memory and their static and dynamic indices. The OpenCL standard requires the compiler to provide built-in functions, that can be called by a kernel. They include functions for organization, synchronization and data access, which are mapped to specific components and control inputs. This representation is also used for hardware-specific optimization, like common subexpression identification. HWMap, the second step in OCLAcc depends on the specific hardware used and thus, exploits vendor-tools. By choosing a supported device, the hardware description is derived. Depending on the vendor and the FPGA, OCLAcc either directly instantiates components, generates IP-cores, or relies on inference by the vendor-tools. Scheduling of components is tightly coupled with their generation, because for many parts of the system, parameters like latency or maximum clock frequency are only available when they have been realized. For this reason, clock-driven synchronization of arithmetic units is only done inside of basic blocks, while among blocks, a simple `ready/ack-protocol` is used.

## IV. NOVELTIES IN OPENCL 2.0

OCLAcc is based on SPIR 2.0 and OpenCL 1.2, though it does not implement the whole standard. This is only necessary to be certified by Khronos and is currently not intended. Instead, we work on features of the recent OpenCL 2.0 standard, of which we expect hardware and developers can benefit from.

*Work-group Functions:* Until OpenCL 2.0, no built-in functions had been available for data transfer between work-items of the same work-group, e.g., broadcast, reduction. Instead, data transfer via local memory and explicit synchronization had been necessary. In most cases, not all work-items can be executed simultaneously, so work-group functions imply synchronization, and can be handled similar to existing functions like `barrier()`. All work-items have to be processed until every item has reached the built-in. Depending on the function, each work-item may provide data used to compute the result for every item. In hardware, basic blocks write their data to local storage, realized by SRAM on the FPGA, and wait for synchronization. When all work-items of the group have reached that point, the result can be computed and directly used as input for the following block in each item.

*Pipes:* Pipes realize packet-based communication channels without explicitly managed indices, in contrast to directly allocating a buffer in memory. Created by the host, pipes are passed to kernels as parameters. Applications with several kernels that consume and produce data, can benefit from pipes as they can directly exchange data without manual synchronization. Their FIFO semantic is well known in hardware design, and their realization using SRAM is obvious, but since size and dimension of a pipe are defined by the host, either generic values have to be used by the hardware implementation or the programmer provides constrains, what we expect to be possible in most cases. Pipes also provide an easy way to exchange data with external devices connected to the FPGA or custom logic on the FPGA, without host interaction, e.g., a data stream from an image sensor is pre- and postprocessed and the results are transmitted to the host via PCIe or directly to any other component connected to the FPGA. This makes pipes the most promising extension in OpenCL 2.0, especially for embedded architectures.

*Device-side enqueue:* Another new feature is device-side enqueue, which allows work-items to launch kernels without having to return to the host. To realize a dynamically spawned function, a hardware implementation has to be available. OCLAcc has to generate them when the main kernels are translated. The work-queue keeps track of all work groups and items scheduled and has been managed solely by the host, but has now to be accessible by the device itself, which is possible by implementing the queue as FIFO that can also be written to by the kernel.

*Shared virtual memory:* Global memory on the device has to be seen as storage of memory objects instead of an address space since addresses are not guaranteed to be preserved across kernel instances or between host and device. By introducing shared virtual memory (SVM), host and kernels may exchange pointers to allocated regions. Three kinds of SVM are introduced, with only the first being required by the standard. Coarse-grained sharing works on buffers which can be mapped and unmapped. Pointers to these areas are valid for host and devices. The two kinds of fine-grained SVM allow sharing on basis of individual memory access or even obviate memory handling by OpenCL and allow kernels and host to use memory allocated by `malloc()`. These kernels have the same address space as the host, e.g. when using the CPU as device. Coarse-grained synchronization is similar to memory management before OpenCL 2.0, with the difference that pointers have to be mapped. Finer grained SVM are not to be implemented by OCLAcc in the near future.

## V. CONCLUSION

This paper gives an overview on how OCLAcc derives hardware descriptions from OpenCL, and discusses new features of OpenCL 2.0, which we expect to ease hardware development or extend the possibilities, without having to extend the standard with non-portable specialties.

## REFERENCES

[1] T. Czajkowski, U. Aydonat, D. Denisenko, J. Freeman, M. Kinsner, D. Neto, J. Wong, P. Yiannacouras, and D. Singh, "From opencl to high-performance hardware on fpgas," in *Field Programmable Logic and Applications (FPL), 2012 22nd International Conference on*, Aug 2012, pp. 531–534.

[2] "Sdaccel development environment," Xilinx, Tech. Rep. UG1023 (v2015.1), Jun 2015. [Online]. Available: http://www.xilinx.com/publications/prod_mktg/sdx/sdaccel-backgrounder.pdf

[3] D. J. Warne, N. A. Kelson, and R. F. Hayward, "Comparison of high level FPGA hardware design for solving tri-diagonal linear systems," *Procedia Computer Science*, vol. 29, pp. 95 – 101, 2014, 2014 International Conference on Computational Science.

[4] M. Owaida, N. Bellas, K. Daloukas, and C. Antonopoulos, "Synthesis of platform architectures from opencl programs," in *Field-Programmable Custom Computing Machines (FCCM), 2011 IEEE 19th Annual International Symposium on*, May 2011, pp. 186–193.

[5] F. Richter-Gottfried and D. Fey, "OCLAcc: An open-source generator for configurable logic block based accelerators," in *Embedded World Conference Proceedings*, Feb 2014.